\documentclass[unsortedaddress,superscriptaddress,prl,twocolumn]{revtex4-2}
\usepackage{hyperref}
\usepackage{epsfig}
\usepackage{graphicx}
\usepackage{subfigure}
\usepackage{latexsym}
\usepackage{color}
\usepackage{fullpage}
\usepackage{dcolumn}
\usepackage{bm}
\usepackage[normalem]{ulem}
\usepackage{units}
\usepackage{amsmath}
\usepackage[paperwidth=210mm,paperheight=297mm,centering,hmargin=2cm,vmargin=2.3cm]{geometry}

\begin{document}

%\title{Impact of the bending energy on the moir\'{e} lattices of two-dimensional materials on substrates}

\title{Elastic properties of moir\'{e} lattices in epitaxial two-dimensional materials}

\author{Alexandre Artaud}
\affiliation{Universit\'{e} Grenoble Alpes, CNRS, Institut NEEL, Grenoble INP, 38000 Grenoble, France}
\affiliation{Universit\'{e} Grenoble Alpes, CEA, Grenoble INP, IRIG, PHELIQS, 38000 Grenoble, France}
\affiliation{Department of Quantum Nanoscience, Kavli Institute of Nanoscience, Delft University of Technology, 2628 CJ Delft, the Netherlands}
\author{Nicolas Rougemaille}
\affiliation{Universit\'{e} Grenoble Alpes, CNRS, Institut NEEL, Grenoble INP, 38000 Grenoble, France}
\author{Sergio Vlaic}
\affiliation{Laboratoire de Physique et d'\'{E}tude des Mat\'{e}riaux, ESPCI Paris, PSL University, CNRS UMR8213, Sorbonne Universit\'{e}s, 75005 Paris, France}
\author{Vincent T. Renard}
\affiliation{Universit\'{e} Grenoble Alpes, CEA, Grenoble INP, IRIG, PHELIQS, 38000 Grenoble, France}
\author{Nicolae Atodiresei}
\affiliation{Peter Gr\"{u}nberg Institute and Institute for Advanced Simulation, Forschungszentrum J\"{u}lich, Wilhelm-Johnen-Stra{\ss}e, 52428 J\"{u}lich, Germany}
\author{Johann Coraux}
\email{johann.coraux@neel.cnrs.fr}
\affiliation{Universit\'{e} Grenoble Alpes, CNRS, Institut NEEL, Grenoble INP, 38000 Grenoble, France}

\begin{abstract}
Unlike conventional two-dimensional (2D) semiconductor superlattices, moir\'{e} patterns in 2D materials are flexible and their electronic, magnetic, optical and mechanical properties depend on their topography. Within a continuous+atomistic theory treating 2D materials as crystalline elastic membranes, we abandon the flat-membrane scenario usually assumed for these materials and address out-of-plane deformations. We confront our predictions to experimental analyses on model systems, epitaxial graphene and MoS$_2$ on metals, and reveal that compression/expansion and bending energies stored in the membrane can compete with adhesion energy, leading to a subtle moir\'{e} wavelength selection and the formation of wrinkles.
\end{abstract}

\maketitle

\begin{figure}[!ht]
\begin{center}
\includegraphics[width=8cm]{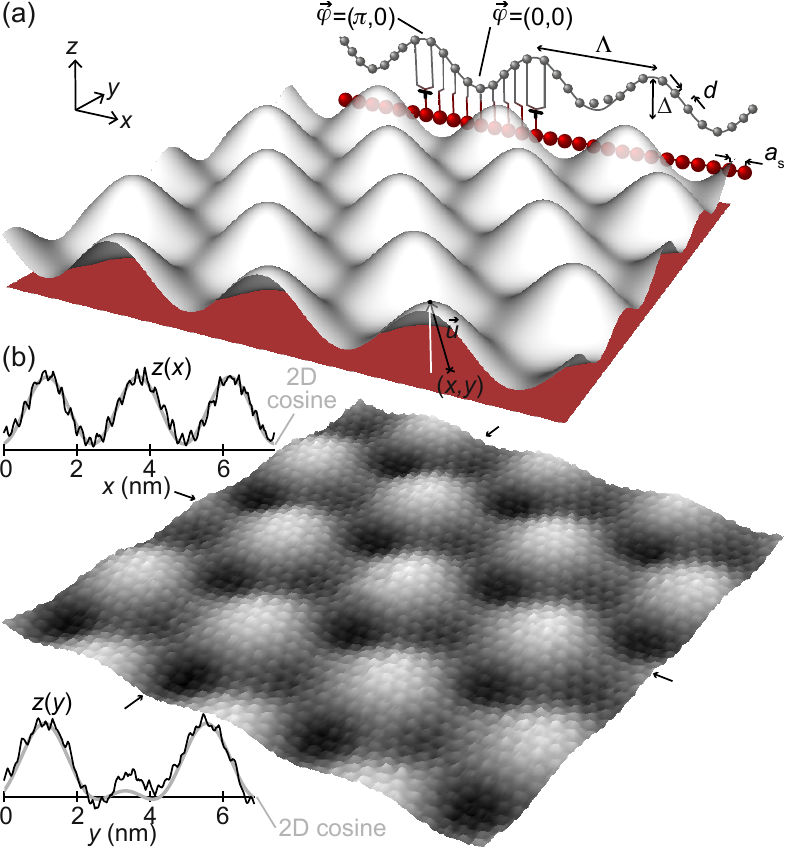}
\caption{\label{fig1} (a) Points on the membrane are displaced from their position on a flat unrippled membrane by $\vec{u}(x,y)$. The cross-section (top-right) shows the moir\'{e} pattern's undulation (period $\Lambda$, amplitude $\Delta$) and the atomic lattices (interatomic distances $d$ and $a_\mathrm{s}$ in the membrane and substrate, respectively). The 2D phase $\vec{\varphi}$, \textit{i.e.} the atomic stacking, varies from a valley to a hill. There, a substrate atom stands halfway between two membrane atoms ($\top$ symbols). (b) STM topograph (8$\times$8~nm$^2$) of a graphene membrane onto Ir(111), and apparent height profiles extracted between the arrows fitted using a cosine function.}
\end{center}
\end{figure}

\textit{\textbf{Introduction. -- }}
Two-dimensional (2D) materials host height fluctuations called nanoripples and are therefore never perfectly flat, behaving as ultimately thin membranes \cite{*Fasolino,*Meyer,*Brivio}. A substrate generally suppresses the dynamics of these height fluctuations \cite{Amorim}. If cystalline, it stabilizes ordered arrays of \textit{static} nanoripples. Their origin lies in the lateral periodic variation of the local atomic stacking (Fig.~\ref{fig1}a), imposed by the lattice mismatch/misorientation of the two materials, and forming a so-called (quasi)coincidence superlattice, \textit{i.e.} a moir\'{e} pattern \cite{Hermann,Artaud,Zeller,Pochet}.

Moir\'{e}s are ubiquitous in epitaxial 2D materials, including graphene \cite{Land}, \textit{h}-BN \cite{Corso}, and MoS$_2$ \cite{Sorensen}. They enrich their electronic properties \cite{Pletikosic,Papagno}, and promote the self-organization of nanoclusters \cite{NDiaye2006,Dil}, molecules \cite{Mao,Dil}, and isolated atoms \cite{Balog,Baltic,Trishin}, with foreseeable unique magnetic and catalytic properties. The mechanical properties are modified too, by phonon localization or phonon branch replicas \cite{Endlich,Maccariello,AlTaleb}, which should manifest in the thermal properties \cite{Amorim}. Rationalizing these properties requires the knowledge of the wavelength $\Lambda$ and amplitude $\Delta$ of the nanoripple pattern (Fig.~\ref{fig1}a). $\Lambda$ is often simply evaluated geometrically, from the superposition of the individual 2D material and substrate lattices. $\Delta$ is often considered to be set by the strength of the interaction with the substrate \cite{Busse,Wang,Decker,Sutter} or a planar stress \cite{Runte}.

Here we investigate, numerically and experimentally, how $\Lambda$ and $\Delta$ are interlinked through the \textit{elasticity} of the 2D membrane. Introducing a mixed continuum mechanics / atomistic modeling we address the usually disregarded effects of non-planar deformations, \textit{i.e.} bending, on nanorippling under the influence of a crystalline surface. This allows to study moir\'{e} systems with numbers of atoms beyond what density functional theory (DFT) and molecular dynamics calculations can tackle, as shown previously for twisted bilayers \cite{Nam,Enaldiev}. We apply our model to two epitaxial 2D materials. Comparing graphene on Ir with and without an intercalated Co monolayer, we investigate the influence of two substrates with the same lattice parameter but different binding strength. Futhermore, we demonstrate the generality of our method with another 2D material, MoS$_2$/Au. Separating the contributions to the total energy, we relate weak nanorippling to a form of the membrane-substrate interaction varying moderately across the moir\'{e}, which generates only small bending energy penalty provided that $\Lambda$ is large enough (graphene/Ir, MoS$_2$/Au). A more subtle $\Lambda$ selection is unveiled when the substrate promotes strong nanorippling (graphene/Co/Ir): the membrane mitigates its bending energy by increasing its planar-projected area. This is accommodated by local wrinkling, as confirmed by microscopy data.

\textit{\textbf{Modeling -- }}
We apply elastic thin plate theory to a membrane having a sinusoidal topography, while taking into account the atomic arrangement at the substrate surface and within the membrane. Two vector fields are considered: the displacement field $\vec{u}$ associated with the membrane deformation and the geometrical phase $\vec{\varphi}\in [0,2\pi]^2$ describing the coincidence between the membrane and substrate atoms (Fig.~\ref{fig1}a). The continuum mechanics and atomistic viewpoints are entangled in $\vec{u}$, which is at the same time a continuous $\vec{u}(x,y)$ and a discrete $\vec{u}_i$ field (defined for each atom $i$ of the membrane). For simplicity, we assume a uniform interatomic distance $d$ over the membrane, and an infinite rigidity of the substrate lattice.

The membrane surface has the form $\Delta/9\sum_{i=1}^3 \cos(\vec{k}_i \cdot \vec{r}) + \Delta/3$ (graphene) and $\Delta/(3\sqrt{3})\sum_{i=1}^3 \sin(\vec{k}_i \cdot \vec{r}) + \Delta/2$ (MoS$_2$) \cite{noteonsurface}, with $\vec{r}$ a 2D position vector and $\vec{k}_{i=1,2,3}$ three $2\pi/3$-rotated vectors of norm $4\pi/(\Lambda\sqrt{3})$. The elastic energy $E_\mathrm{el}=1/2\int\mathrm{Tr}\hspace{1pt}(\varepsilon\sigma)dxdy=\int{e_\mathrm{el}}dxdy$ is decomposed, using the stress-strain ($\sigma$ and $\varepsilon$ tensors) relationship and the Lam\'{e} coefficients ($\lambda$, $\mu$), bending rigidity ($\kappa$), and Poisson ratio ($\nu$) \cite{Landau}, in in-plane (ip) and out-of-plane (oop) components \cite{noteonuz}:

\begin{equation}
\begin{split}
E_\mathrm{el,ip} =& \frac{1}{2} \int \big(\lambda (\varepsilon_{xx}+\varepsilon_{yy})^2 \\
&+ 2 \mu (\varepsilon_{xx}^2+\varepsilon_{yy}^2+2\varepsilon_{xy}^2)\big) dxdy \\
E_\mathrm{el,oop} =& \frac{\kappa}{2} \int \big( (\partial_{x}\partial_{x} u_z
+ \partial_y\partial_y u_z)^2 + 2(1 - \nu)\\ 
&\times((\partial_x\partial_y u_z)^2 - \partial_x\partial_x u_z \partial_y\partial_y u_z  ) \big) dxdy
\end{split}
\label{eq1}
\end{equation}

\noindent with $\varepsilon_{\alpha\beta}=1/2(\partial_\beta u_\alpha + \partial_\alpha u_\beta + \sum_{\tau=x,y,z} \partial_\alpha u_\tau \partial_\beta u_\tau )$, $\alpha,\beta=x,y$, and $\partial_{\alpha,\beta}=\partial/\partial_{\alpha,\beta}$.

The adhesion energy of the membrane on the substrate writes as a sum over the atomic positions:

\begin{equation}
E_\mathrm{ad} = \sum_i e_{\mathrm{ad},i}(\vec{\varphi}_i,u_{z,i})
\end{equation}

\begin{table*}[!ht]
\caption{\label{tab:energies} Calculated elastic $\bar{e}_\mathrm{el}$ and adhesion $\bar{e}_{\mathrm{ad}}$ contributions to the total energy $\bar{e}_\mathrm{t}$, normalized by the number of atoms (meV/\AA$^2$), in a moir\'{e} unit cell, for optimal values of $d$ variation (\%), $\Lambda$ (\AA), and $\Delta$ (\AA). $\Lambda$=$\Lambda'$ and $\Delta$=$\Delta'$ values minimize $\bar{e}_{\mathrm{ad}}$ alone. Experimental structural parameters from the literature and our STM measurements ($^*$) are reported.}
\begin{ruledtabular}
\begin{tabular}{lllllllll}
   & $\Delta d$ & $\Lambda$ & $\Delta$ & $\bar{e}_{\mathrm{ad}}$ & $\bar{e}_\mathrm{el}$ & $\bar{e}_\mathrm{t}$ & $\Lambda'$ & $\Delta'$ \\

\hline
   Graphene/Ir & -0.05 & 26.4 & 0.39 & -11.94 & 0.04 & -11.90 & 26.4 & 0.43 \\

   Experiments & -0.01/-0.29 \cite{Blanc} & 25.5, 25.6 \cite{Hattab,Blanc}, & 0.6/1.0, 0.42/0.56, 0.38 &  &  &  &  &  \\
	&  & 25.4$^*$ & \cite{Busse,Hamalainen,Jean}, 0.35$^*$ &  &  &  &  &  \\

   Graphene/Co/Ir & +0.17 & 27.3 & 1.67 & -12.56 & 0.85 & -11.71 & 25.3 & 2.03 \\

   Experiments & +0.1/1.4 \cite{Gargiani} & 26.5/28.5 \cite{Gargiani} & 1.2/1.8 \cite{Decker}, 1.8$^*$ &  &  &  &  &  \\
   MoS$_2$/Au & -0.25 & 35.2 & 0.44 & -30.74 & 0.23 & -30.50 & 32.8 & 0.46 \\
   Experiments & -0.32, +0.13 \cite{Bana,Silva} & 33.4, 33.3 \cite{Sant,Silva} & 0.37 \cite{Silva} &  &  &  &  &  \\

\end{tabular}
\end{ruledtabular}
\end{table*}

The elastic and adhesion energies are functions of $\Lambda$, $\Delta$, the relative orientation $\theta$ of the membrane and substrate lattices (Sec.~S1.7 of the Supplemental Material \cite{SM}), and $d$ \cite{noteonddep}. Calculating $E_\mathrm{ad}$ requires knowledge of the $e_{\mathrm{ad},i}$ potential, which depends on the kind of substrate and the local membrane-substrate atomic coincidence. For epitaxial 2D materials, $e_{\mathrm{ad},i}$ has at least one minimum, for heights of the 2D material that change within the moir\'{e} cell (with $i$). This promotes nanorippling, hence an $e_\mathrm{el}$ penalty (unless $d$ is compressed) that tends to mitigate it. We search for the lowest-energy structure of the membrane, among the set of $E_\mathrm{el}+E_\mathrm{ad}$ values calculated over a unit cell of the nanoripple pattern, for an extended range of $\lbrace d,\Lambda,\Delta\rbrace$ triplets. We also assessed the influence of $\theta$.

For each system, 6,000 triplets were used, varying $d$ within $\pm$1\% around graphene's or MoS$_2$'s reference values (2.462~\AA, 3.167~\AA; Sec.~S1.2 of the Supplemental Material \cite{SM}), $\Lambda$ across 20-30~\AA\; or 28-38~\AA\; (graphene, MoS$_2$), and $\Delta$ across 0.05-2.4~\AA\; -- by steps of 0.25\%, 0.2~\AA, and 0.1~\AA\;, respectively. Based on DFT calculations, including our own new ones for MoS$_2$/Au(111), accounting for van der Waals interactions at the membrane/substrate interface, we parametrized the adhesion potentials. The elastic constants were taken from the relevant calculations and experimental estimates (Secs.~S1.1,S1.3,S1.4 of the Supplemental Material \cite{SM}). The minimum-energy configurations were then compared to our scanning tunneling microscopy (STM) measurements (Sec.~S2 of the Supplemental Material \cite{SM}), and other previously published experimental data.

\textit{\textbf{Weak nanorippling. -- }}
High-resolution measurement of $\Lambda$, $\Delta$ and $d$ is notoriously challenging experimentally. Graphene on Ir(111) is one of the few systems for which this has been done \cite{Busse,Hattab,Blanc,Jean,Hamalainen,Runte}, and MoS$_2$/Au(111) another one, albeit to a lesser extent \cite{Bana,Sant,Silva}. The measured structural parameters are reported in Table~\ref{tab:energies} -- $\Delta$ estimates vary with the tunneling imaging conditions for graphene/Ir \cite{Hamalainen}, even more so for MoS$_2$/Au \cite{Silva}. A typical STM topograph of graphene/Ir(111) is shown in Fig.~\ref{fig1}b. Apparent height profiles through the moir\'{e} pattern, along high-symmetry directions, are well described by the 2D cosine function introduced above (Fig.~\ref{fig1}b).

A Morse potential faithfully describes the adhesion energy, $e_{\mathrm{Ir-C},i}$, in particular the presence of a large-distance energy minimum ($>$3~\AA). This minimum only slightly varies with the local atomic coincidence; it is much related to a van der Waals interaction that prevails at weak-interaction interfaces between graphene and metals like Cu, Ag, Ir, Pt, Au \cite{Olsen,Christian}. The potential has been parametrized to obtain an average binding energy per C atom close to the 50~meV value derived from DFT calculations (Sec.~S1.3 of the Supplemental Material \cite{SM}). Computing the elastic energy requires knowledge on the elastic constants $\lambda$, $\mu$, $\nu$, which have been estimated for graphene/Ir(111) \cite{Politano}, whereas $\kappa$ is only known for graphite \cite{Nicklow} (Sec.~S1.1 of the Supplemental Material \cite{SM}).

The calculated minimum-energy $d$ and $\Delta$ (Table~\ref{tab:energies}) fit within the range of experimental values \cite{Busse,Jean,Hamalainen}, whereas $\Lambda$ is slightly larger (we will come back to that). Interestingly, the elastic energy marginally contributes to the total energy. Considering the adhesion energy alone leads to similar estimates of $\Lambda$ and $\Delta$ (noted $\Lambda'$ and $\Delta'$ in Table~\ref{tab:energies}, Figs.~S6b,c of the Supplemental Material \cite{SM}). The topography is essentially inherited from the C-Ir interaction ($\Lambda$ and $\Delta$ Table~\ref{tab:energies}), and graphene/Ir(111) is a weakly nanorippled system storing few elastic energy. This holds too for graphene slightly twisted (fractions of degrees are often observed experimentally \cite{Ndiaye2008,Hamalainen}) with respect to Ir(111) (Sec.~S1.7 of the Supplemental Material \cite{SM}).

\begin{figure}[!ht]
\begin{center}
\includegraphics[width=8cm]{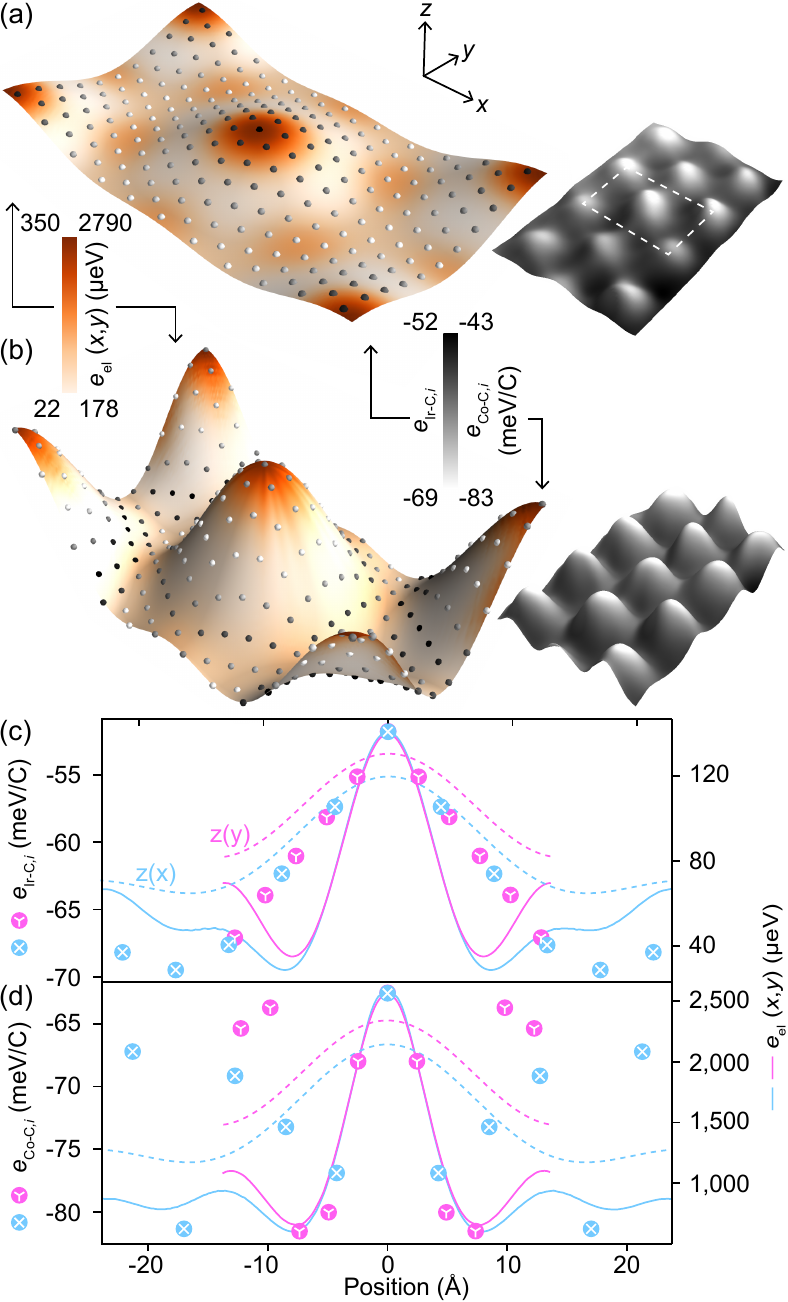}
\caption{\label{fig3} (a,b) Spatial distribution of the elastic ($e_\mathrm{el}$ evaluated on a square grid with 0.2~\AA\; steps) and adhesion ($e_{\mathrm{Ir-C},i}$, $e_{\mathrm{Co-C},i}$) energy densities for graphene on Ir and Co/Ir, and corresponding STM topographs (9$\times$6.5~nm$^2$, one unit cell sketched with a dotted frame). The $z$ scale is multiplied by 10. (c,d) Cross-sections, along $x$ and $y$ axes, of the membrane's height (dotted lines) and corresponding elastic (solid lines) and adhesion (`X' and `Y' symbols) energies, for graphene/Ir (c) and graphene/Co/Ir (d).}
\end{center}
\end{figure}

As the elastic energy reflects the membrane curvature, it is inhomogeneous in space, with $E_\mathrm{el,oop}$ (Eq.~\ref{eq1}) as its main contribution. It is maximum at the top of the nanoripples (dark regions in Fig.~\ref{fig3}a), minimum along their flanks (white regions), and takes intermediate values at the surface's saddle points and valleys (orange regions).

The spatial distribution of the adhesion energy is simpler (Fig.~\ref{fig3}a,c): it follows the surface topography, with the weakest (resp. strongest) binding at the hills (resp. valleys). This reflects the varying graphene-on-Ir stacking configuration, with half the C atoms located on top of the Ir ones (valleys), and the center of C hexagons on top of Ir atoms (hills) \cite{Busse}.

For MoS$_2$/Au, we calculate three times stronger adhesion energy, and six times stronger elastic energy than for graphene/Ir (Table~\ref{tab:energies}). Their spatial distribution is also well explained by the varying local stacking and membrane bending (Fig.~S5 of the Supplemental Material \cite{SM}). Although it is more costly to bend MoS$_2$ than graphene, the energy penalty still appears affordable, presumably owing to the large $\Lambda$ value.

\textit{\textbf{Strong bending effects. -- }}
To gain insight on the influence of adhesion on nanorippling we now consider graphene on Ir with Co intercalated (Sec.~S2 of the Supplemental Material \cite{SM}). The Co surface is pseudomorphic to Ir (same lattice constant) but has a different kind of adhesion. Compared to graphene/Ir, a five-to-ten-fold increase of $\Delta$ is found in STM, depending on the tunnel bias voltage (Figs.~\ref{fig3}a,b) \cite{Decker,Pacile}. 

In the discussion above, the adhesion energy $e_{\mathrm{Ir},i}$ had a single minimum, at a (large) distance varying only slightly with the atomic coincidence, which promoted weak nanorippling. According to \textit{ab initio} calculations \cite{Olsen,Christian}, surfaces having high affinity with C, especially Co and Ni, lead to the occurence of a second minimum at shorter distance ($\sim$2.2~\AA) for coincidences with half C atoms atop a metal atom (Figs.~S1,S3, Sec.~S1.3 of the Supplemental Material \cite{SM}). Adhesion energy variations are in the same range as for $e_{\mathrm{Ir},i}$, but the large distance between the two minima promotes strong nanorippling. Our analytical form of $e_{\mathrm{Co},i}$ adds a Gaussian component to the Morse potential to account for the two minima.

The $\lbrace d,\Lambda,\Delta\rbrace$ triplet minimizing the system energy yields a four times larger $\Delta$ value than for graphene/Ir and agrees well with the experimental estimates (Table~\ref{tab:energies}, Refs.~\citenum{Decker,Pacile}). $\Lambda$ is also substantially larger in the presence of the intercalated Co layer (which is pseudomorphic to Ir). Our calculations show (Fig.~S7b of the Supplemental Material \cite{SM}) that this results from the six-times stronger out-of-plane (bending) energy density here. Note the larger (by about 0.2\%) interatomic distance $d$, now significantly off the reference value. The corresponding in-plane (stretching) elastic energy penalty, more than ten times that in graphene/Ir is compensated by a gain in adhesion energy, allowed by the larger $d$ that yields favorable substrate-membrane atomic coincidences (while the opposite effect is associated with the increased rippling). These behaviours are also found for small twist angles of the graphene lattice (Sec.~S1.7 of the Supplemental Material \cite{SM}).
 
As expected, the spatial distribution of the elastic energy density is essentially the same as for graphene/Ir (Figs.~\ref{fig3}b,d), most of the energy being stored at high curvature regions, \textit{i.e.} the hills, the valleys and the saddle points. The spatial distribution of the adhesion energy density is more complex here. In particular, the valleys of the membrane are not anymore the only regions with strong adhesion energy, and the spatial variations are much faster (Figs.~\ref{fig3}c,d). Changing the metal substrate thus deeply modifies the adhesion energy density within the membrane.

\textit{\textbf{Mechanical instability. -- }}
Our calculations predict $\Lambda$ values larger by $\sim$1~\AA\; and $\sim$2~\AA, respectively, than experimental values for graphene/Ir \cite{Hattab,Blanc} and MoS$_2$/Au \cite{Silva,Sant} (Table~\ref{tab:energies}). Such discrepancies are expected as a result of the heteroepitaxial stress building up as the samples are cooled down after growth, due to the mismatch in thermal expansion coefficients of the 2D material and the substrate. In graphene/Ir this stress is only partially relieved to the expense of a local bending and loss of adhesion, by linear delaminations called wrinkles \cite{Hattab}; no wrinkles form to relieve the (smaller) stress in MoS$_2$/Au \cite{Bana}. In other words, experimentally graphene/Ir and MoS$_2$/Au are close but not exactly at the calculated equilibrium state. For their lower $\Lambda$ values, we calculate lower-estimates of the excess total energy, compared to the equilibrium state, of only 4~meV (graphene/Ir) and 2~meV (MoS$_2$/Au) per unit cell. This is because the total energy weakly depends on $\Lambda$ when $\Delta\simeq$ 0.4~\AA\; (Fig.~S6c of the Supplemental Material \cite{SM}). If thermodynamic equilibrium is not precisely reached experimentally (growth is an out-of-equilibrium process), the total energy is only weakly affected.

This is not the case with graphene/Co/Ir, for which this energy difference is, depending on the graphene twist angle (0$^\circ$, 0.5$^\circ$, 1$^\circ$), 20 to 60 times larger due to strong elastic energy variations at large $\Delta$ values (Fig.~S6d of the Supplemental Material \cite{SM}). As discussed above this is the driving force for a complex $\Delta$ and $\Lambda$ selection, beyond what the adhesion energy alone would impose. This large excess bending energy, in graphene/Co/Ir with $\Delta$ $\simeq$ 25.5~\AA, may be relieved by an increase of $\Lambda$, hence an increase of the projected membrane's area. The latter must be accommodated somewhere. This is what experiments reveal: a second network of wrinkles forms upon Co intercalation. This new network is easily recognized: wrinkles are lower, shorter, and denser than those formed after graphene growth (Fig.~S8 of the Supplemental Material \cite{SM}).

\begin{figure}
\begin{center}
\includegraphics[width=8cm]{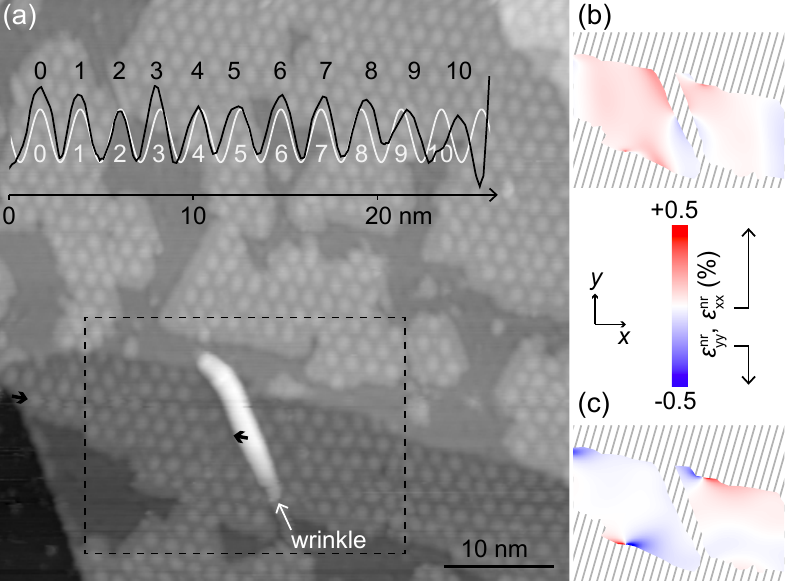}
\caption{\label{fig4} (a) STM topograph of graphene/Ir intercalated with a Co submonolayer. A graphene wrinkle appears as a bright linear feature. Inset: apparent height profile between the two black arrows, compared to a sinusoid (the successive maxima are numbered). (b,c) Maps of the moir\'{e} lattice compression/expansion in the $xy$ plane, roughly perpendicular (c) and parallel (d) to the wrinkle.}
\end{center}
\end{figure}

One wrinkle in this new network, at an intermediate stage of the intercalation process, is visible in Fig.~\ref{fig4}a. We analyzed the in-plane deformations $\varepsilon_{xx}^{\mathrm{nr}}$ and $\varepsilon_{yy}^{\mathrm{nr}}$ in the $x$ and $y$ directions of the nanoripple (nr) lattice at the vicinity of the wrinkle using a geometrical phase analysis (Sec.~S3 of the Supplemental Material \cite{SM}). A gradient of $\varepsilon_{xx}^{\mathrm{nr}}$ is observed perpendicular to the wrinkle (Fig.~\ref{fig4}b): the nanoripple lattice is stretched when approaching the wrinkle. In fact, the moir\'e lattice expands by several tenths of percent, bringing $\Lambda$ to values close to, and even at, those we predict for the lowest-energy configuration (Table~\ref{tab:energies}). Consistently, in (apparent) height profiles, $\Lambda$ increases as the distance to the wrinkle shortens (Fig.~\ref{fig4}a). On the contrary, in the direction parallel to the wrinkle there is no obvious $\varepsilon_{yy}^{\mathrm{nr}}$ variation close to the wrinkle (Fig.~\ref{fig4}c). Other effects seem to play a more important role in this direction, for instance the presence of a substrate step edge or the edge of the intercalated Co island.

Altogether, our spatially-resolved analysis of the deformation field in the nanoripple pattern supports the above proposal for a wrinkling mechanism induced by an `unrippling' ($\Lambda$ increase) of the membrane. This mechanism highlights the far-reaching consequences of the mechanical (elastic) back-action of the membrane under the influence of a substrate.

\textit{\textbf{Prospects. -- }}
Our model predicts the ground-state structure of graphene and MoS$_2$ on substrates with different adhesion properties. Experimental data are close to these predictions, with only minor discrepancies that vanish if bending energy penalties matter, \textit{i.e.} for strongly rippled systems. We identify where and how the 2D membrane stores energy \textit{via} elastic deformations, bending in particular, and disentangle the role of the substrate from that of the membrane's mechanical properties. Altogether these are of utmost importance to understand and engineer nanorippling-related properties, \textit{e.g.} pseudo-electromagnetic fields, excitonics, electronic correlations \cite{Dai}. Our approach is complementary to first-principles ones, permitting fast calculations ($<$ 1~s) on large systems and the exploration of a broad range of parameters ($\Lambda$, $a_\mathrm{s}$, $\theta$). It is also applicable to twisted 2D bilayers in the presence of strain fields \cite{Nam,Enaldiev} -- altogether, to help interpret the rich moir\'{e}-related phenomena, including disorder \cite{Blanc,Hamalainen,Ochoa,Koshino} and temperature effects.

\begin{acknowledgements}
\textit{\textbf{Acknowledgements. -- }}We thank P. David and V. Guisset for valuable support with the ultrahigh vacuum experiments and J.-L. Rouvi\`{e}re for providing a geometrical phase analysis software. N.A. acknowledges Deutsche Forschungsgemeinschaft (DFG) support through the Collaborative Research Center SFB 1238 Project number 277146847 (subproject C01). The authors gratefully acknowledge the computing time granted by the JARA Vergabegremium and provided on the JARA Partition part of the supercomputer JURECA at Forschungszentrum J\"{u}lich. V.T.R acknowledges the suport from ANR Flatmoi project (ANR-21-CE30-0029). We thank Andrea Locatelli, Tevfik Onur Mente\c{s} and Benito Santos for assistance during the low-energy electron microscopy measurements.
\end{acknowledgements}

%\bibliography{2D_moire_undulations_ripples}

\begin{thebibliography}{72}
\bibitem [{not({\natexlab{b}})}]{noteonuz}%
  \BibitemOpen
  \href@noop {} {} \bibinfo {note} {Note that $u_z$
  is not simply the height on membrane surface, but the $z$ coordinate of the
  elementary surface element on the membrane, displaced by the deformation of
  the latter (Fig.~\ref{fig1}a).}\BibitemShut {Stop}%
\bibitem [{not({\natexlab{c}})}]{noteonddep}%
  \BibitemOpen
  \href@noop {} {}  \bibinfo {note} {the latter dependency is
  explicit for $E_\mathrm{ad}$ and implicit for $E_\mathrm{el}$, corresponding
  to a change of the total membrane surface when $d$ deviates from the
  reference unstrained value.}\BibitemShut {Stop}%
\end{thebibliography}

%apsrev4-2.bst 2019-01-14 (MD) hand-edited version of apsrev4-1.bst
%Control: key (0)
%Control: author (8) initials jnrlst
%Control: editor formatted (1) identically to author
%Control: production of article title (0) allowed
%Control: page (0) single
%Control: year (1) truncated
%Control: production of eprint (0) enabled
%

\end{document}